УДК: 2013.12.27

# Возбуждение бабблов и бризеров в ДНК и их взаимодействие с носителями заряда

©2014 Лахно В.Д.[*1], Четвериков А.П.[**2]

[1]*Институт математических проблем биологии, Российская академия наук, Пущино, Московская область, 142290, Россия*

[2]*Саратовский государственный университет им. Чернышевского, Саратов, 410012, Россия*

***Аннотация.*** На основе модели Пейрарда–Бишопа исследованы нелинейные возбуждения в ДНК, вызванные внешним воздействием. Показано, что такое воздействие может приводить к появлению бабблов, способных распространяться вдоль молекулы. Рассмотрено взаимодействие бабблов с помещённым в цепочку избыточным зарядом: электроном или дыркой. Проведены численные эксперименты, демонстрирующие возможность переноса заряда бабблами. Обсуждается возможность применения полученных результатов в нанобиоэлектронике.

***Ключевые слова:*** *модель Пейрарда–Бишопа–Доксуа, нанобиоэлектроника, подвижность, полярон.*

ВВЕДЕНИЕ

В связи с развитием молекулярной нанобиоэлектроники (НБЭ), основной задачей которой является конструирование электронных устройств на основе биологических молекул [1, 2], все больший интерес вызывают проблемы транспорта заряда в протяженных биомолекулах. Решение этой и других задач, обеспечивающих использование биомолекул в молекулярной электронике, позволило бы решить многочисленные проблемы, возникающие в кремниевой наноэлектронике, такие как миниатюризация, самосборка, самокопирование и другие.

В настоящее время в качестве базовой биомолекулы, на использовании свойств которой будет базироваться будущая нанобиоэлектроника, рассматривается молекула ДНК. В первую очередь это обусловлено открытой в 90-х годах прошлого столетия способностью ДНК проводить электрический ток [3], см. также обзор [4]. Другая основополагающая причина, по которой ДНК является наиболее перспективной молекулой с точки зрения ее использования в НБЭ, состоит в том, что она является единственной молекулой, способной продуцировать саму себя. Из молекул ДНК можно создавать сложные структуры и электрические цепи для конструирования различных НБЭ устройств. В настоящее время не составляет труда синтезировать ДНК с любой заданной длиной последовательностью (до ~ $10^3$) нуклеотидных пар. Эти и многие другие свойства молекулы ДНК делают ее уникальным претендентом на использование в создании элементной базы будущей нанобиоэлектроники.

Созданные на основе ДНК нанопроводники должны обладать возможностью эффективно переносить заряд на большое расстояние. В качестве движущей силы такого переноса обычно рассматривают разность потенциалов, создаваемую между концами нанопровода. Однако всё больший интерес вызывают не только эффекты переноса заряда в ДНК в электрическом поле, но и другие возможные механизмы транспорта заряда. Согласно [5, 6], теоретические оценки подвижности заряда в ДНК дают малое значение этой величины $\mu \sim 1 \div 10^{-4}$ см$^2$/В·сек. Так, из известных данных расчётов для Poly G/Poly C цепочек, где G обозначает гуанин, C – цитозин, следует, что в «сухой» цепочке величина дырочной подвижности $\mu$ составляет порядка 1 см$^2$/В·сек [7–9]. Для «мокрых» цепочек, т. е. цепочек в растворе, где большую роль играет эффект сольватации, подвижность оказывается на несколько порядков ниже, чем в «сухих» цепочках [5, 6]. Использование проводов из ДНК с такой малой подвижностью делает их малоперспективными для целей нанобиоэлектроники. Хорошо известно, однако, что молекула ДНК сама по себе имеет возможность переносить нелинейные локализованные возбуждения даже в отсутствие введенных в нее зарядов. К числу таких возбуждений относятся, например, кинки и солитоны [10]. Исключительно большой интерес представляют также такие нелинейные возбуждения в ДНК как пузыри, ответственные за денатурацию молекулы и разделение цепей в процессе транскрипции [11–14]. Они могут распространяться вдоль молекулы ДНК с большой скоростью и при введении в молекулу избыточных зарядов могут с ними взаимодействовать. В частности, известно описание образования связанных состояний возбужденных в ДНК солитонов с введенными в нее электронами. В работах [15–17] такие связанные состояния были названы солектронами.

---

[*]lak@impb.psn.ru
[**]chetvap@rambler.ru

Целью данной работы является исследование инициированных внешним воздействием нелинейных возбуждений ДНК в рамках модели Пейрарда–Бишопа (ПБ) [18–21] и их взаимодействия с внесенным в ДНК избыточным зарядом (электроном или дыркой). Отметим, что в настоящее время модель ПБ, а также ее модификация – модель Пейрарда–Бишопа–Доксуа (ПБД), – наилучшим образом описывает динамические и статистические свойства ДНК, включая такие явления как образование в ней пузырьков (бабблов) и плавление молекулы (см. обзор [22]).

Статья организована следующим образом.

В разделе 2 сначала приведены гамильтониан системы и самосогласованные уравнения, описывающие полуклассическую динамику входящих в состав молекулы ДНК нуклеотидов, рассматриваемых как классические частицы, и взаимодействующего с ними квантового электрона (дырки), эволюция волновой функции которого подчиняется уравнению Шредингера.

Затем в разделе 3 предполагается, что избыточный заряд в молекуле ДНК отсутствует и посредством численного моделирования изучается эволюция линейных и нелинейных волн, возбужденных за счет возмущений скорости одного или нескольких сайтов ДНК, вызываемых внешним импульсным или гармоническим воздействием.

В разделе 4 мы анализируем взаимодействие таких возбуждений с избыточным зарядом, первоначально локализованным на одном или небольшом числе нуклеотидных пар. Рассматриваются свойства связанных состояний, образованных бризером и захваченным им электроном, а также связанных состояний солитонного типа, движущихся с высокой скоростью.

В разделе 5 излагаются результаты моделирования движения поляронов, инициированное возбужденными в молекулярной цепочке бризерами и бабблами.

В разделе 6 подводятся итоги, обсуждаются физические аспекты обнаруженных эффектов, намечаются пути дальнейших исследований.

## МОДЕЛЬ

Будем проводить исследование свойств транспорта электрона в ДНК в рамках квантово-классической модели, являющейся комбинацией модели Холстейна [23] и Пейрарда–Бишопа [18], имея в виду, что простое описание динамики возмущений в цепочке сайтов, принятое в модели Холстейна, заменено на более точное, предложенное в работе [18]. В частности, для описания взаимодействия нуклеотидов в паре используется нелинейный потенциал Морзе, а сами пары, которые при моделировании будем называть сайтами, рассматриваются как связанные линейными силами звенья квази-одномерной цепочки (усовершенствованный вариант модели цепочки, в котором учитывается, например, и нелинейность сил взаимодействия сайтов [24, 22] предполагается исследовать отдельно). Гамильтониан модели в этом случае может быть представлен в виде

$$\hat{H} = \sum_n^N \alpha_n |n\rangle\langle n| + \sum_{n,m}^N v_{nm} |n\rangle\langle n| + \chi \sum_n (w_n - v_n)|n\rangle\langle n| +$$
$$+ \sum_n \left[ \frac{1}{2} m(\dot{w}_n^2 + \dot{v}_n^2) + \frac{k}{2}((w_n - w_{n-1})^2 + (v_n - v_{n-1})^2) + U(w_n - v_n) \right], \quad (1)$$

где $\alpha_n$ – энергия заряда на *n*-ном сайте, $v_{nm}$ – коэффициенты «перескоковых» интегралов, $\chi$ – константа связи внешнего заряда и цепочки, $w_n$, $v_n$ – смещения отдельных нуклеотидов в *n*-ой нуклеотидной паре от положения равновесия, точка означает производную по времени, а соответствующие величины – скорости нуклеотидов в *n*-ой нуклеотидной паре, *m* – масса нуклеотида. Входящая в (1) величина $U(w_n - v_n)$ определяет взаимодействие нуклеотидов в паре, средний член в квадратной скобке в (1) определяет взаимодействие между соседними вдоль цепи нуклеотидами (стэкинговое взаимодействие).

Если искать решение, соответствующее гамильтониану (1), в виде

$$|\Psi\rangle = \sum_n c_n(t)|n\rangle \quad (2)$$

и ввести новые переменные

$$x_n = (w_n + v_n)/\sqrt{2}, \qquad y_n = (w_n - v_n)/\sqrt{2}, \quad (3)$$

то соответствующие уравнения движения нуклеотидных пар и дискретное уравнение Шредингера для электрона выглядят следующим образом

$$m\frac{d^2 y_n}{dt^2} + \kappa(2y_n - y_{n-1} - y_{n+1}) + \frac{\partial}{\partial y_n}U(y_n) + \sqrt{2}\chi|c_n|^2 = 0, \quad (4)$$

$$i\hbar \frac{dc_n}{dt} = \alpha_n c_n + \sum_m v_{nm} c_m + \sqrt{2}\chi y_n c_n. \quad (5)$$

Отсюда для polyG/polyC однородной цепочки в приближении ближайших соседей получим $(\alpha_n \equiv 0)$

$$i\hbar \frac{dc_n}{dt} = -v(c_{n+1} + c_{n-1}) + \sqrt{2}\chi y_n c_n. \quad (6)$$

Таким образом, переписывая уравнения с учетом введенных переменных, получим

$$\frac{d^2 y_n}{dt^2} + \frac{\kappa}{m}(2y_n - y_{n-1} - y_{n+1}) + \frac{1}{m}\frac{\partial}{\partial y_n}U(y_n) + \frac{\sqrt{2}\chi}{m}|c_n|^2 = 0, \quad (7)$$

$$\frac{dc_n}{dt} = i\frac{v}{\hbar}(c_{n+1} + c_{n-1}) - i\frac{\sqrt{2}\chi}{\hbar}y_n c_n. \quad (8)$$

Полагая, что потенциал взаимодействия двух нуклеотидов сайта, находящихся на расстоянии *r* между ними, может быть аппроксимирован потенциалом Морзе [18,19]

$$U(y_n) = D(e^{-2\sigma y_n} - 2^{-\sigma y_n}), \quad (9)$$

представим силу взаимодействия между нуклеотидами в паре в форме:

$$F = -\frac{\partial}{\partial r}U(y_n) = 2\sigma D(e^{-\sigma y_n} - 1)e^{-\sigma y_n}. \quad (10)$$

Здесь коэффициент упругости σ характеризует жесткость связи. Введем переменные, ассоциированные с взаимодействием в каждой нуклеотидной паре: $q_n = \sigma y_n$ – нормированное смещение, $\tau = \omega_M t$ – безразмерное время, где $\omega_M$ – частота линейных колебаний в изолированном сайте ($\omega_M^2 = 2\sigma^2 D/m$), $\omega_{bond}^2 = \kappa/2\sigma^2 D$ – нормированная на $\omega_M$ частота колебаний сайтов в цепочке, $\chi_h = \sqrt{2}\chi/2\sigma D$ и $\chi_{el} = \sqrt{2}\chi/\hbar\omega_M\sigma = \chi_h(2D/\hbar\omega_M)$ безразмерные параметры связи электрона и решетки, $\tau_e = \nu/\hbar\omega_M$ – безразмерный параметр, задающий отношение характерных времен эволюции волновой функции электрона и динамики возмущений в решетке, (это означает, что безразмерные единичные пространственные и временные масштабы в «безразмерной» модели будут определяться динамикой частиц в сайте, а межсайтовые взаимодействия будут задавать эволюцию возмущений с характерными пространственными масштабами, существенно большими единицы), в результате будем иметь

$$\ddot{q}_n = e^{-q_n}(e^{-q_n} - 1) + \omega_{bond}^2(q_{n+1} - 2q_n + q_{n-1}) - \chi_h|c_n|^2, \tag{11}$$

$$\dot{c}_n = -i\tau_e(c_{n+1} + c_{n-1}) - i\chi_{el}q_n c_n. \tag{12}$$

В линейном приближении по $q$ и при $\chi_h = 0$ из (11) следует

$$\ddot{q}_n = -q_n + \omega_{bond}^2(q_{n+1} - 2q_n + q_{n-1}), \tag{13}$$

и известное дисперсионное уравнение

$$\omega^2 = 1 + 4\omega_{bond}^2 \sin^2\left(\frac{k}{2}\right), \tag{14}$$

откуда вытекает, что фононный спектр заключен между частотами 1 и $\sqrt{1 + 4\omega_{bond}^2}$. Здесь $k$ – безразмерное волновое число в масштабе, определяемом величиной σ.

Уравнения (11), (12) со специфическим заданием начальных условий (см. ниже) и периодическими граничными условиями решались численно методом Рунге–Кутты 4-го порядка с обеспечением необходимой точности, в частности, поддержанием постоянными значений интегралов системы (прежде всего, энергии рассматриваемой консервативной системы и вероятности найти внешнюю частицу в системе, равной 1). Характерный шаг интегрирования выбирался 0.001 или 0.0005, число сайтов $N = 50$–200, характерные времена симуляций $10^2$–$10^3$. Однако прежде чем перейти к описанию и анализу результатов симуляций, оценим типичные значения безразмерных параметров модели, исходя из известных значений физических величин, определяющих эти параметры. Будем полагать, что $D = 0.04$ эВ, $\sigma = 4.45$ Å$^{-1}$, $\kappa = 0.06$ эВ/Å [19], $m = 300$ а.е.м. (а.е.м.=$1.66 \cdot 10^{-27}$ кг – атомная единица массы), $\chi = 0.13$ эВ/Å, $\nu = 0.084$ эВ [25] (для Poly G/Poly C - цепочки), $\hbar = 6.5822 \cdot 10^{-16}$ эВ·сек – постоянная Планка. Отсюда следует что $\omega_{bond} = \sqrt{K/2a^2D} \approx 0.2$, $\tau_e = \nu/\hbar\omega_M \approx 18$, $\chi_h \approx 0.5$, $(2D/\hbar\omega_M) \approx 17$ и тогда $\chi_{el} \approx 8.5$. Значения параметров, используемые в численном моделировании, выбирались с учетом проведенных оценок.

Особенностью данного исследования является способ возбуждения колебаний и волн в цепочке. Предполагается, что необходимая для этого энергия вводится посредством возмущения скорости одного или нескольких смежных сайтов внешним локальным воздействием, причем время воздействия достаточно короткое. В результате в цепочке сайтов должны возникать нелинейные локализованные волны, которые будут в состоянии захватывать электрон и перемещать его вдоль цепочки. Отметим, что возмущение решетки за счет начальных *смещений* частиц из положений равновесия рассматривалось, например, в работе [26]. Было показано, что в этом случае возможно формирование нелинейных локализованных волн в форме бризеров. Мы, напротив, будем рассматривать начальное возмущение *скоростей* частиц ДНК, т.е. предполагать, что в систему вводится начальная кинетическая энергия. (Мы не будем сейчас описывать экспериментальные реализации такого способа, анализируя пока его принципиальные возможности).

## НЕЛИНЕЙНЫЕ КОЛЕБАНИЯ И ВОЛНЫ В ДНК, ВОЗБУЖДАЕМЫЕ ВНЕШНИМ ВОЗМУЩЕНИЕМ СКОРОСТИ САЙТОВ

Рассмотрим несколько способов возмущения скорости частиц недеформированной цепочки внешним воздействием, задавая различные начальные условия для уравнения движения, вытекающего из (11) при $\chi_h = 0$

$$\ddot{q}_n = e^{-q_n}\left(e^{-q_n} - 1\right) + \omega_{bond}^2 \left(q_{n+1} - 2q_n + q_{n-1}\right) \quad (16)$$

Во-первых, если считать, что в начальный момент в цепочке один из элементов получает начальный импульс $p_{in} = mv_{in}$, то в слабо-нелинейном режиме (при относительно небольшой величине импульса) наблюдается формирование неподвижного бризера с частотой $\omega_{br}$, меньшей критической частоты (рис. 1, $v_{in} = 1$, $\omega_{br} \approx 0.7$). Компоненты спектра начального импульса, принадлежащие фононному спектру, при этом рассеиваются в решетке. Однако если импульс большой (рис. 2, $v_{in} = 16$), то образуется баббл с симметричными резкими фронтами, которые затем начинают медленно распространяться в форме бризеро-подобных возмущений (именно в них, как показано ниже, может захватываться электрон). Колебания в середине баббла со временем ослабляются, и он окончательно разрывается и превращается в два баббла-бризера, движущихся в противоположных направлениях. Максимальная разница смещений в проведенном эксперименте даже при очень большом уровне «закачиваемой» мгновенно в решетку энергии (например, безразмерная скорость $v_{in} = 16$ (рис. 2) соответствует кинетической энергии 256D, поскольку энергия измеряется в используемой модели в единицах 2D, где, напомним, D – глубина ямы потенциала Морзе (9)) не превышает 20/σ ~5 Ангстрем. Это представляется допустимой величиной, с учетом величины диаметра молекулы ДНК, оцениваемой как ~ 20 ангстрем.

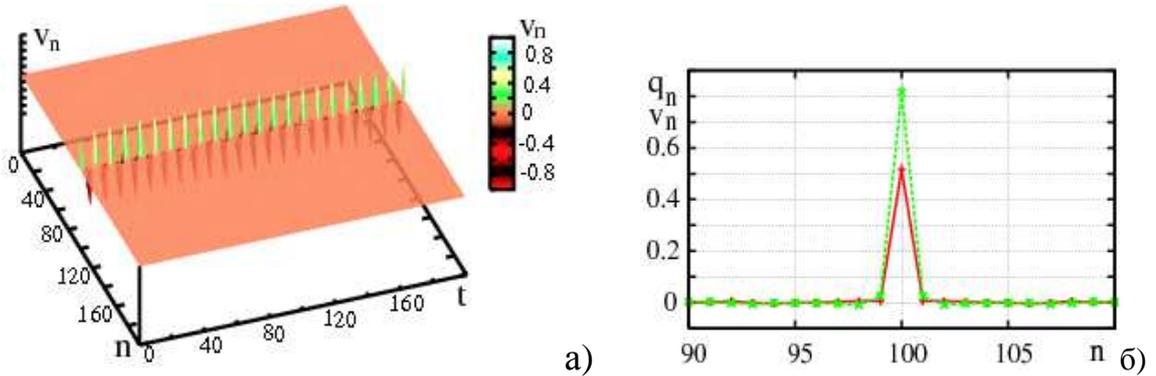

**Рис. 1.** Формирование неподвижного бризера при возбуждении цепочки начальным импульсом одной частицы ($v_{in} = 1$, $n_{in} = 100$, $N = 200$). Слева, а), показана эволюция во времени распределения скорости частиц, справа, б), фрагмент распределения смещений $q_n$ (красным) и скоростей $v_n$ (зеленым) частиц в завершающий момент симуляций. Частота бризера $\omega_{br} \approx 0.7$.

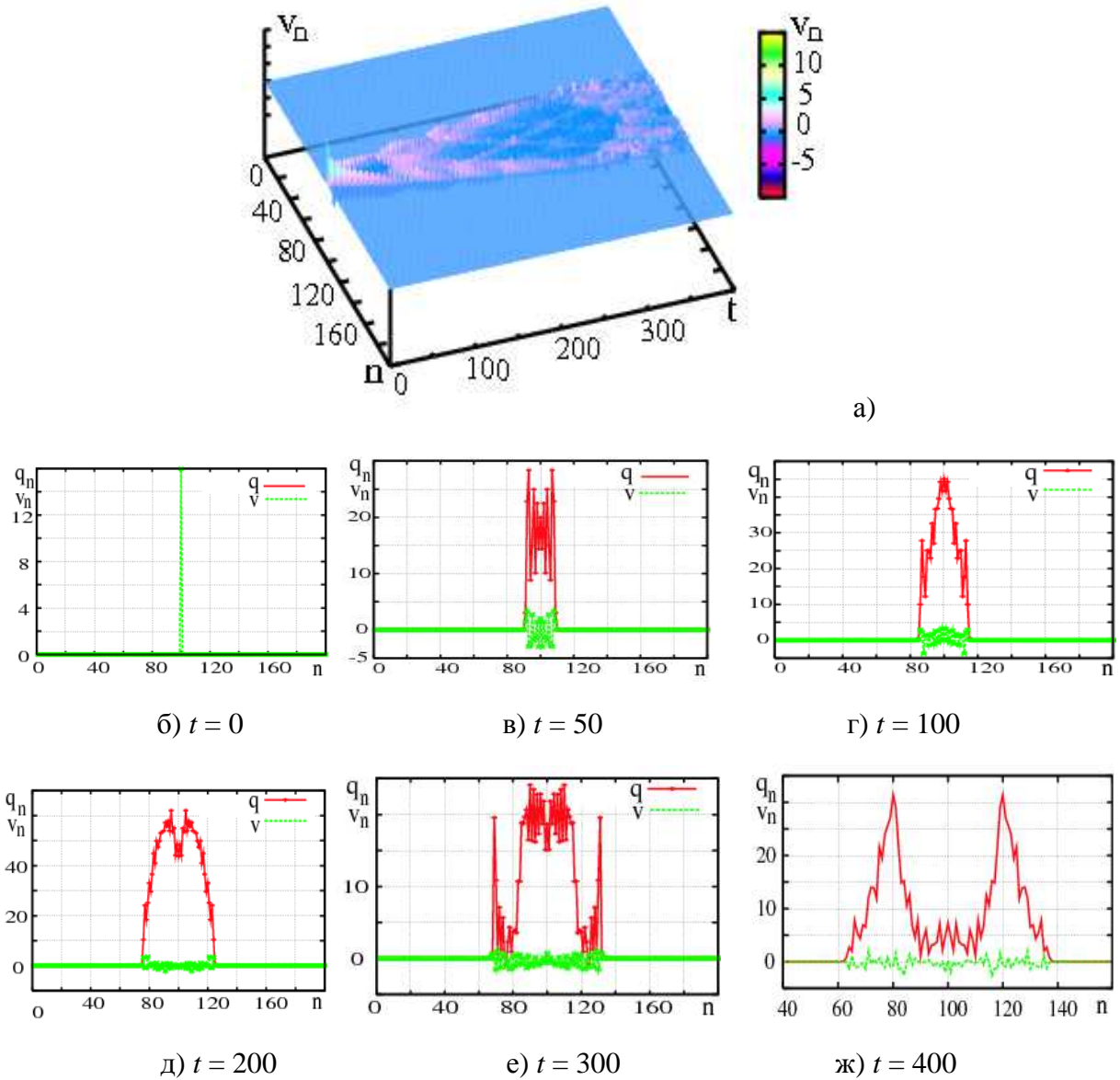

**Рис. 2.** Возбуждение цепочки сильным начальным импульсом одной частицы ($v_{in} = 16$, $n_{in} = 100$, $N = 200$). Показаны эволюция во времени распределения скорости частиц, а), и распределения смещений $q_n$ (красным) и скоростей $v_n$ (зеленым) частиц в нескольких последовательных моментах времени, б)–ж).

Во-вторых, перспективно реализовать возбуждение молекулы внешней периодической силой (накачкой), действующей на один или несколько смежных сайтов. В этом случае существует принципиальная возможность варьировать значения амплитуды и частоты накачки, время ее действия, число возбуждаемых элементов и разность фаз колебаний между осцилляторами. Пусть сначала на один осциллятор с номером $n_{in}$ действует гармоническая сила, возмущающая скорость частицы, т. е. добавляющая в скорость компоненту $v(n_{in}) = v_{in}\cos(\omega_{in}\tau)$ в течение примерно одного периода колебаний возбуждающей силы, $\Delta\tau \sim T_{in} = 2\pi/\omega_{in}$. Величина $v_{in}$ выбирается достаточно большой, чтобы за короткое время закачать в решетку много энергии. Частота $\omega_{in}$ выбирается меньшей 1 (т. е. меньше безразмерной $\omega_M$ – нижней частоты полосы фононного спектра), чтобы внешняя сила была в среднем в резонансе с частотой осциллятора, которая уменьшается с ростом энергии. В этом случае опять образуется баббл с симметричными резкими фронтами, которые затем начинают распространяться в форме бризеров, которые могут захватывать электрон. В целом картина оказывается похожа на представленную выше (сравни рис. 2,а и рис. 3).

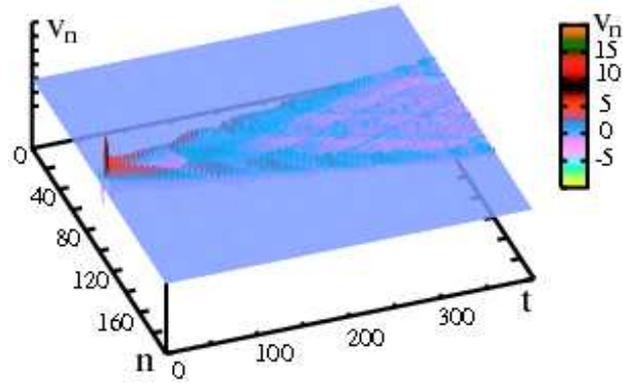

**Рис. 3.** Возбуждение цепочки периодическим (гармоническим) возмущением скорости одной частицы. Показана эволюция во времени скоростей частиц $v_n(\tau)$. $n_{in} = 100$, $N = 200$, $v_{in} = 0.002$, $\omega_{in} = 0.7$, $\Delta\tau = 10$.

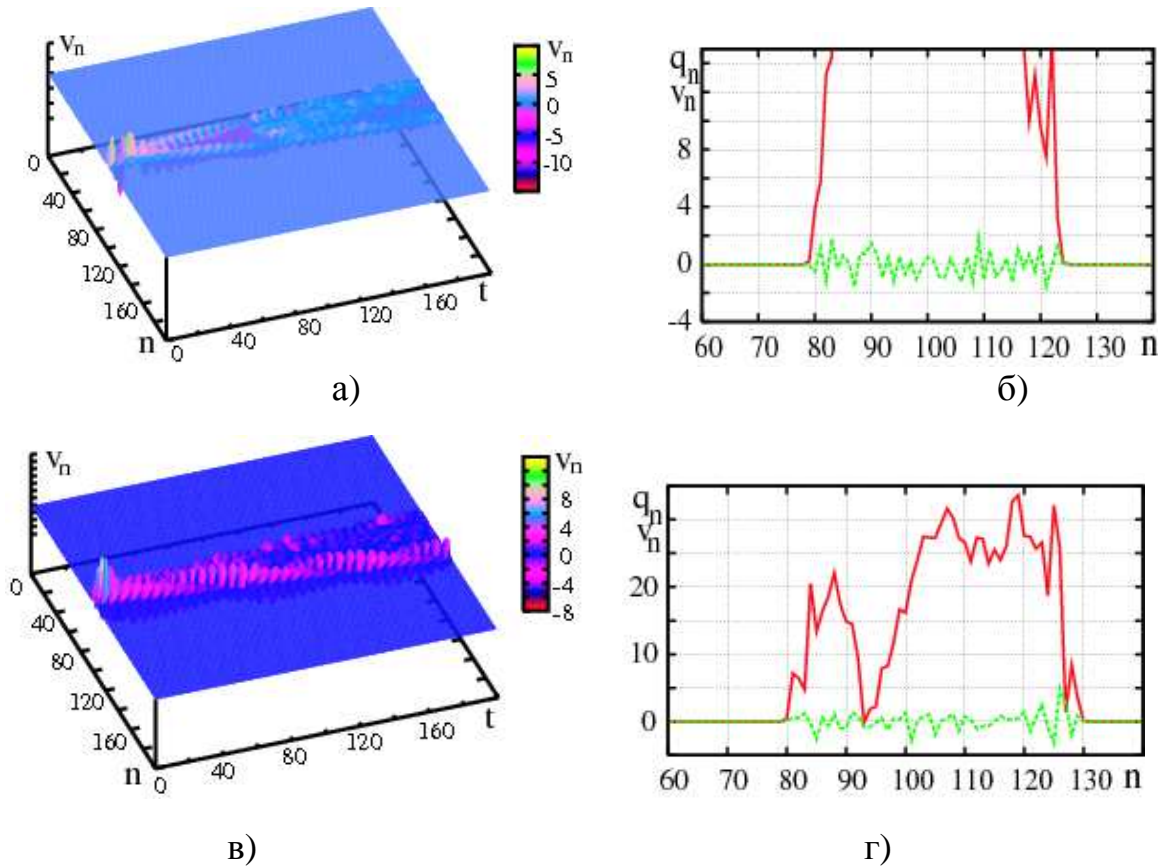

**Рис. 4.** Возбуждение цепочки периодическим (гармоническим) возмущением скорости двух частиц с номерами $n_{in1} = 100{-}101$ (а, б) и четырех частиц с номерами $n_{in1} = 100{-}103$ (в, г) цепочки из $N = 200$ сайтов. Показаны эволюция во времени скоростей частиц $v_n(\tau)$ (а, в) и фрагменты распределений смещений $q_n$ (красным) и скоростей $v_n$ (зеленым) частиц в завершающий момент симуляции (б, г) $v_{in} = 0.002$, $\omega_{in} = 0.7$, $\Delta\tau = 10$, $\Delta\varphi = -1.2$.

В-третьих, чтобы обеспечить образование связанного состояния с преимущественным направлением движения в одну сторону, нужно возбуждать, по крайней мере, два сайта со сдвигом фаз колебаний $\Delta\varphi$ возмущающего периодического воздействия (на рис. 4,а и 4,б видно, что бризер на правом фронте баббла более развитый, чем на левом). Однако эффект несимметричности формирования бризеров в этом случае выражен слабо и для того чтобы усилить его, нужно использовать возбуждение нескольких смежных сайтов со сдвигом фаз колебаний (рис. 4,в и 4,г), подобно тому, как в электронике возбуждают, например, поверхностно-акустическую волну излучателем-гребенкой [27]. В частности, из рисунков 4,в и 4,г на которых представлены результаты возбуждения цепочки за счет внешней накачки четырех смежных сайтов, видно, что бризер на фронте баббла, распространяющийся при той же заданной разности фаз внешних колебаний $\Delta\varphi = -1.2$ вправо, гораздо более мощный, и основная энергия возбуждения сосредоточена именно в нем. Если же задать положительную разность фаз, то основной бризер будет распространяться влево (здесь не показано).

Можно также полагать, что разность фаз возбуждения можно реализовать не только между соседними сайтами, но и между группами сайтов (кластерами), в каждом из которых частицы колеблются в фазе. Отметим, что исследование динамики ДНК, находящейся под внешним воздействием, представляет самостоятельный интерес, в частности, стимулированный быстрым развитием терагерцовых технологий [28], однако в рамках настоящей работе мы ограничиваемся лишь демонстрацией принципиальных возможностей такого способа возбуждения молекулы ДНК.

## ВЗАИМОДЕЙСТВИЕ ЭЛЕКТРОНА С НЕЛИНЕЙНЫМИ ЛОКАЛИЗОВАННЫМИ ВОЛНАМИ В ДНК

Рассмотрим теперь взаимодействие сформированных локализованных возмущений в решетке (бабблов и бризеров) с вносимым в нее избыточным зарядом. Будем сначала для простоты полагать, что в момент внешнего удара электрон локализован на том сайте невозмущенной решетки, который подвергается внешнему воздействию. На рисунках 5–8 представлены картины эволюции волновой функции электрона (точнее, дискретной функции распределения $|c_n|^2$ вероятности обнаружить электрон на $n$-ом сайте) и ее состояние, а также состояние распределений смещений и скоростей частиц в последний момент симуляции в случае возбуждения решетки способами, описанными выше.

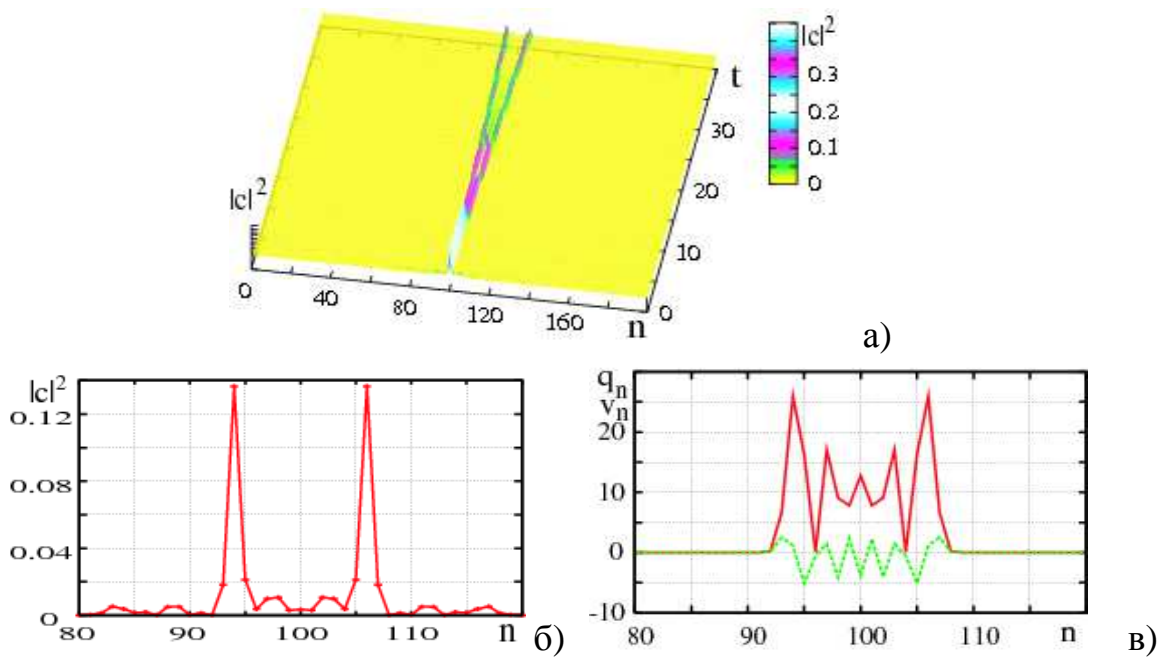

**Рис. 5.** Возбуждение баббла и бризеров начальным импульсом одной частицы ($v_{in} = 16$, $n_{in} = 100$, $N = 200$) и захват электрона ($\omega_{bound} \approx 0.2$, $\tau_e = 18$, $\chi_h = 0.5$, $\chi_{el} = 6$). Показаны эволюция во времени распределения плотности вероятности обнаружить электрон в цепочке, а), распределение плотности вероятности (б) и смещений $q_n$ (красным) и скоростей $v_n$ (зеленым) частиц (в) вдоль цепочки в последний момент симуляции при $\tau = 40$.

Из рис. 5, демонстрирующего результаты захвата электрона, оказавшегося в начальный момент времени (без образования полярона) на сайте, который получает сильный внешний импульс как на рис. 2, видно, что волновая функция электрона локализуется на фронтах расширяющегося баббла, частицы в каждом из которых организованы в форме бризера. В результате происходит транспорт электрона со скоростью бризера. Однако расстояние, на которое смещается электрон (примерно 6 в каждую сторону за время 40) не очень большое, поскольку примерно в момент времени $\tau = 60$ связанное состояние разрушается - бризер теряет электрон и волновая функция «размазывается» вдоль решетки (здесь не показано). Аналогичные картины можно наблюдать и в случае возбуждения бризера в решетке периодическим возмущением скорости одного сайта (рис. 6). В этом случае электрон смещается на расстояние примерно 12 сайтов (в каждую сторону) за время 80, разрушается связанное состояния примерно в момент времени $\tau = 90$. Чтобы обеспечить образование связанного состояния с преимущественным направлением движения в одну сторону, как уже отмечалось выше, нужно возбуждать, по крайней мере, два сайта со сдвигом фаз колебаний возмущающего периодического воздействия (см. рис. 4). При этом селекция бризеров не является очень эффективной, они отличаются не очень сильно (рис. 4,а и 4,б), однако электрон захватывается преимущественно, как показывают результаты моделирования, только в один из бризеров, причем время жизни связанного состояния возрастает (рис. 7).

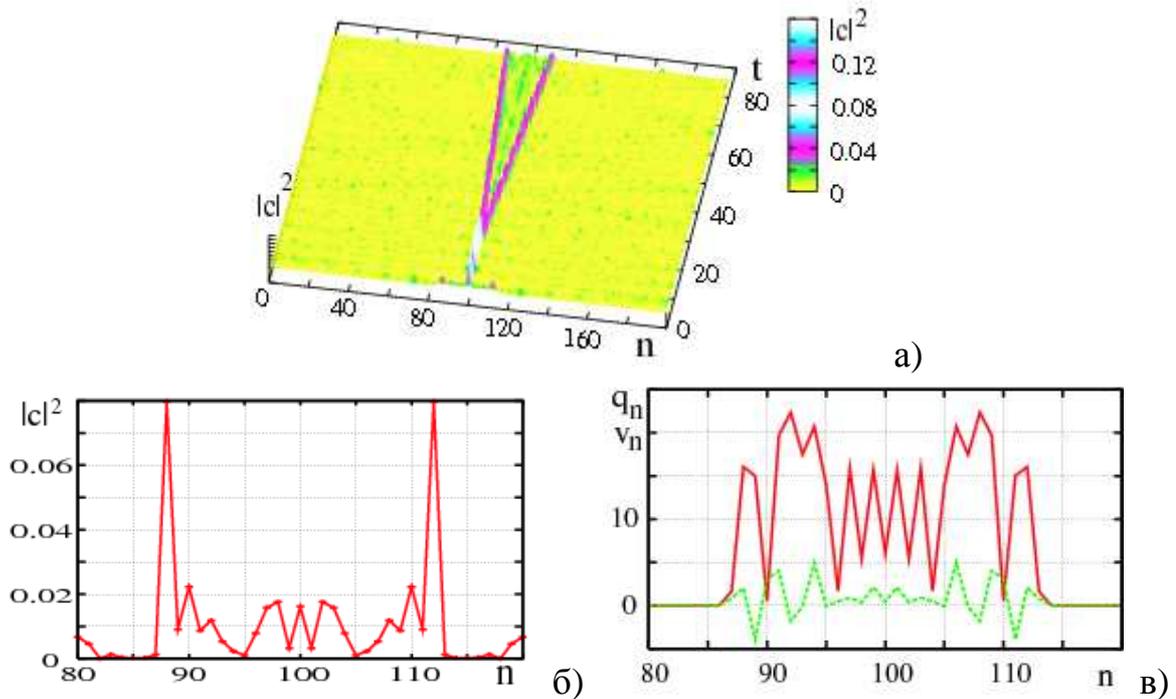

**Рис. 6.** Возбуждение баббла и бризеров периодическим внешним возмущением скорости одной частицы и захват электрона. Показаны эволюция во времени распределения плотности вероятности обнаружить электрон в цепочке, а), распределение плотности вероятности (б) и смещений $q_n$ (красным) и скоростей $v_n$ (зеленым) частиц (в) вдоль цепочки в последний момент симуляции при $\tau = 80$, $N = 200$, $n_{in} = 100$, $v_{in} \approx 0.002$, $\Delta\tau = 10$. Остальные параметры имеют значения, указанные выше ($\omega_{bound} \approx 0.2$, $\tau_e = 18$, $\chi_h = 0.5$, $\chi_{el} = 6$; $\omega_{in} \approx 0.7$).

Эффект усиливается с возрастанием количества возбуждаемых сайтов. В частности, при возбуждении 4 сайтов можно обеспечить очень высокую селекцию возбуждения «полярона-бризера» (polarobreather, [29]), (рис. 8). Это связано с тем, что при возбуждении двух сайтов наблюдается преимущественное направление движения связанного объекта, но бризеры в решетке почти симметричны. А чтобы обеспечить сразу заданное направление движения образующегося бризера, нужно использовать возбуждение нескольких смежных (а может быть, и не обязательно смежных) сайтов со сдвигом фаз возбуждающих колебаний.

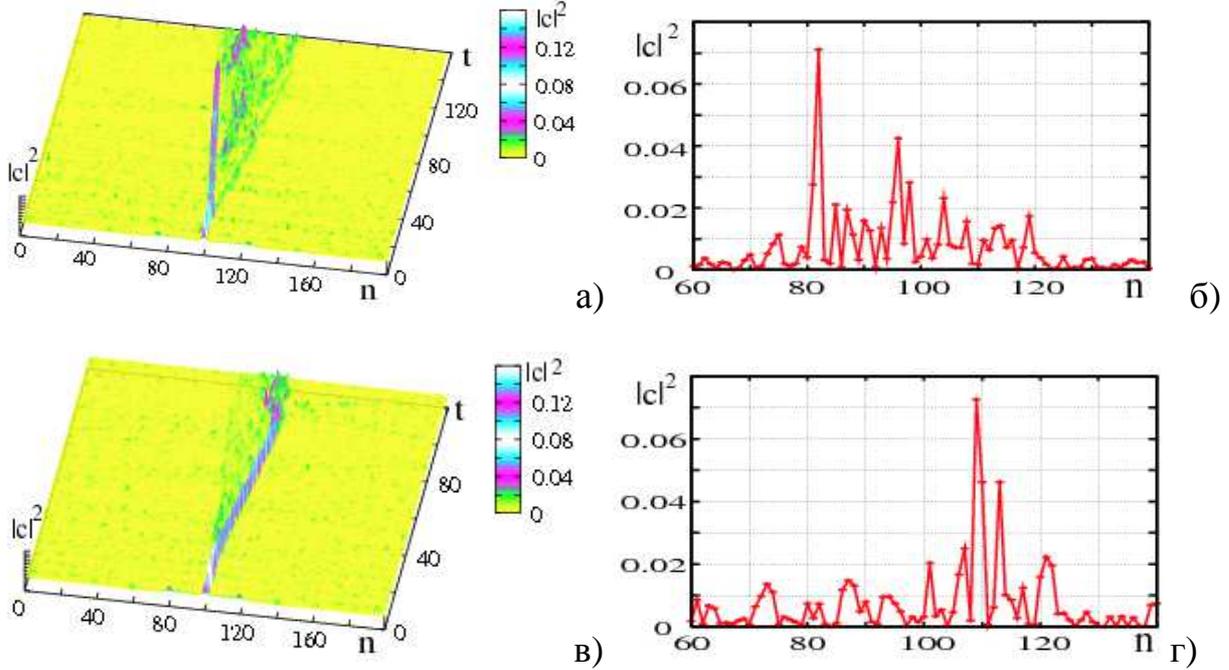

**Рис. 7.** Возбуждение баббла и бризеров периодическим внешним возмущением скоростей двух частиц ($n_{in}$ = 100–101, $v_{in} \approx 0.002$, $\Delta\tau = 10$) и захват электрона для двух значений разности фаз возмущающего воздействия, $\Delta\varphi = 1$ (а, б) и $\Delta\varphi = -1.4$ (в, г). Показаны эволюция во времени распределения плотности вероятности обнаружить электрон в цепочке, а) и в), и распределение плотности вероятности вдоль цепочки, б) и г), в момент времени перед разрушением связанного состояния ($\tau = 120$ для а) и $\tau = 100$ в)). Остальные параметры имеют значения, указанные выше ($N = 200$, $\omega_{bound} \approx 0.2$, $\tau_e = 18$, $\chi_h = 0.5$, $\chi_{el} = 6$; $\omega_{in} \approx 0.7$).

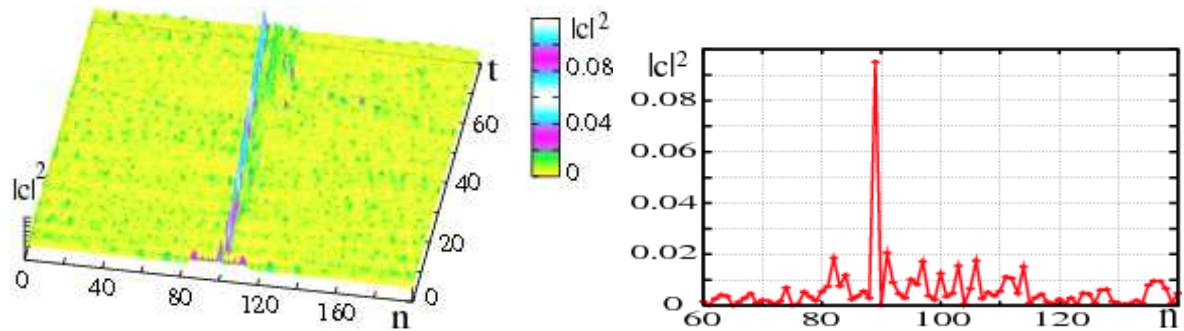

**Рис. 8.** Возбуждение баббла периодическим внешним возмущением скоростей четырех частиц ($n_{in}$ = 100–103, $v_{in} \approx 0.002$, $\Delta\tau = 10$, $\Delta\varphi = 1$) и захват электрона. Показаны эволюция во времени распределения плотности вероятности обнаружить электрон в цепочке (слева) и фрагмент распределения плотности вероятности вдоль цепочки (справа) в момент времени перед разрушением связанного состояния ($\tau = 80$). $N = 200$, $\omega_{bound} \approx 0.2$, $\tau_e = 18$, $\chi_h = 0.5$, $\chi_{el} = 6$; $\omega_{in} \approx 0.7$.

Завершая раздел, отметим, что варьированием параметров в области разрешенных значений можно пытаться найти комплекс значений, при которых время жизни связанного состояния будет максимальным, но мы оставляем эту задачу на будущее.

## ВЗАИМОДЕЙСТВИЕ ПОЛЯРОНА С БАББЛАМИ В ДНК

Предположим теперь, что локализованный первоначально в невозмущенной решетке неподвижный электрон имеет возможность сформировать полярон и лишь затем в решетке возбуждается баббл одним из описанных способов. Стационарное поляронное состояние электрона в ДНК хорошо изучено и возможность его существования в рассматриваемой системе сомнения не вызывает (см., например, [30, 31] и соответствующие ссылки). Отметим, однако, что при изучении эволюции поляронного состояния необходимо учитывать, что характерные времена эволюции существенно зависят от параметра взаимодействия $\chi_h$. В частности, при используемом выше значении параметра $\chi_h = 0.5$ времена эволюции очень большие. Даже при $\chi_h = 5$ характерное время эволюции достаточно большое – об этом можно судить по представленной на рис. 9 картине эволюции во времени распределения плотности вероятности обнаружить электрон $|c_n|^2$ в случае, когда первоначально локализованная на сайте с номером $n = 100$ функция распределения сначала быстро делокализуется вдоль невозмущенной решетки и лишь спустя длительное время вновь локализуется, образуя полярон. Отметим, что в поляроне молекула не расширяется в поперечном сечении как в баббле, а наоборот сжимается ($q_n < 0$, рис. 9,в), а энергия полярона значительно меньше, чем энергия рассмотренных выше связанных состояний на бризере, в которых скорость частиц решетки на порядок выше. Поэтому полярон чувствителен ко всяким возмущениям в решетке, в частности, влиянию фононов. Для того чтобы в какой-то мере предотвратить влияние фононов, возникающих в описанном процессе формирования полярона, в численном моделировании в уравнение движения (11) дополнительно вводилась диссипация, характеризуемая коэффициентом $\gamma$ с типичными используемыми значениями в диапазоне 0.001–0.1. Используя затем данные о состоянии решетки и волновой функции электрона в последний момент моделирования (рис. 9,б и 9,в), можно наблюдать процесс формирования полярона и в молекуле с меньшими значениями параметра $\chi_h$, протекающий сравнительно быстро (рис. 10). Отметим, что благодаря влиянию периодических граничных условий полярон формируется преимущественно либо вблизи области начальной локализации электрона (в рассматриваемом случае – в середине решетки вблизи сайта с номером $n = 100$ решетки с $N = 200$ сайтами, рис. 10,а и 10,б), либо со сдвигом на N/2 сайтов (в нашем случае - на границе ячейки моделирования, рис. 10,в–е).

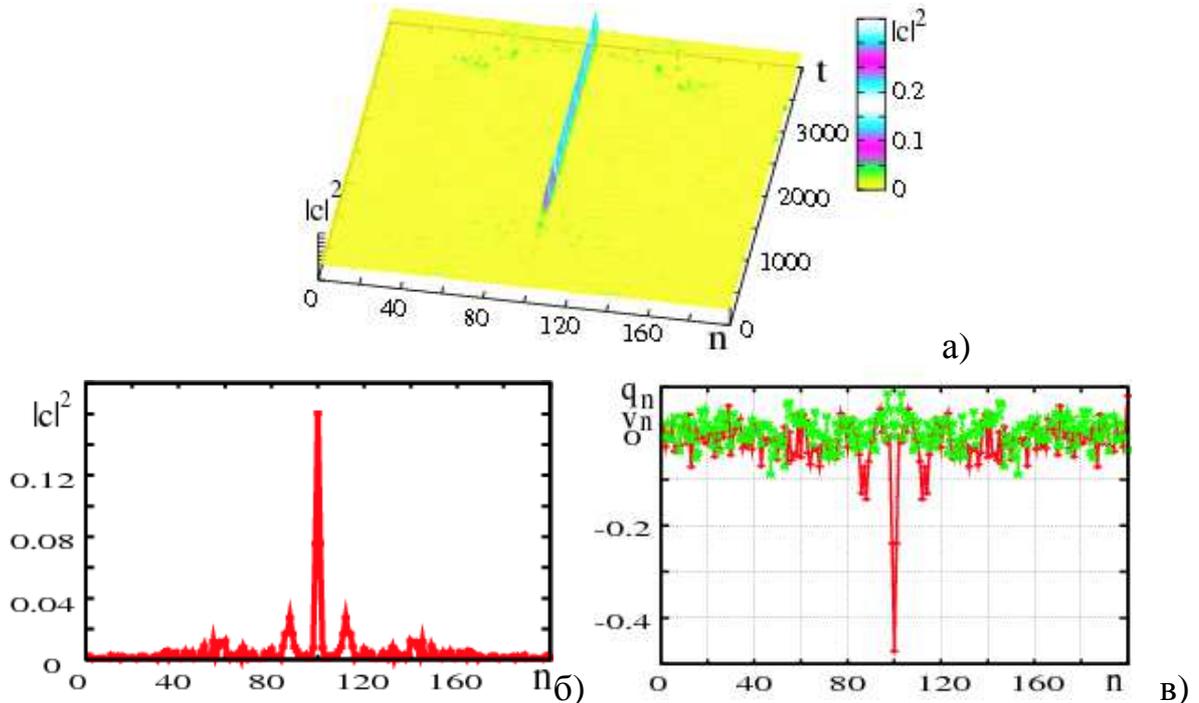

**Рис. 9.** Формирование полярона первоначально локализованным на одном сайте ($n = 100$) электроном. Представлены картина эволюции во времени распределения плотности вероятности обнаружить электрон в цепочке, а), распределение плотности вероятности обнаружить электрон на $n$-ом сайте, б), и смещений $q_n$ (красным) и скоростей $v_n$ (зеленым) частиц вдоль цепочки, в), в последний момент симуляции при $\tau = 4000$. $N = 200$, $\chi_h=5$, $\tau_e = 18$, $\chi_{el} = 85$, коэффициент диссипации $\gamma = 0.001$.

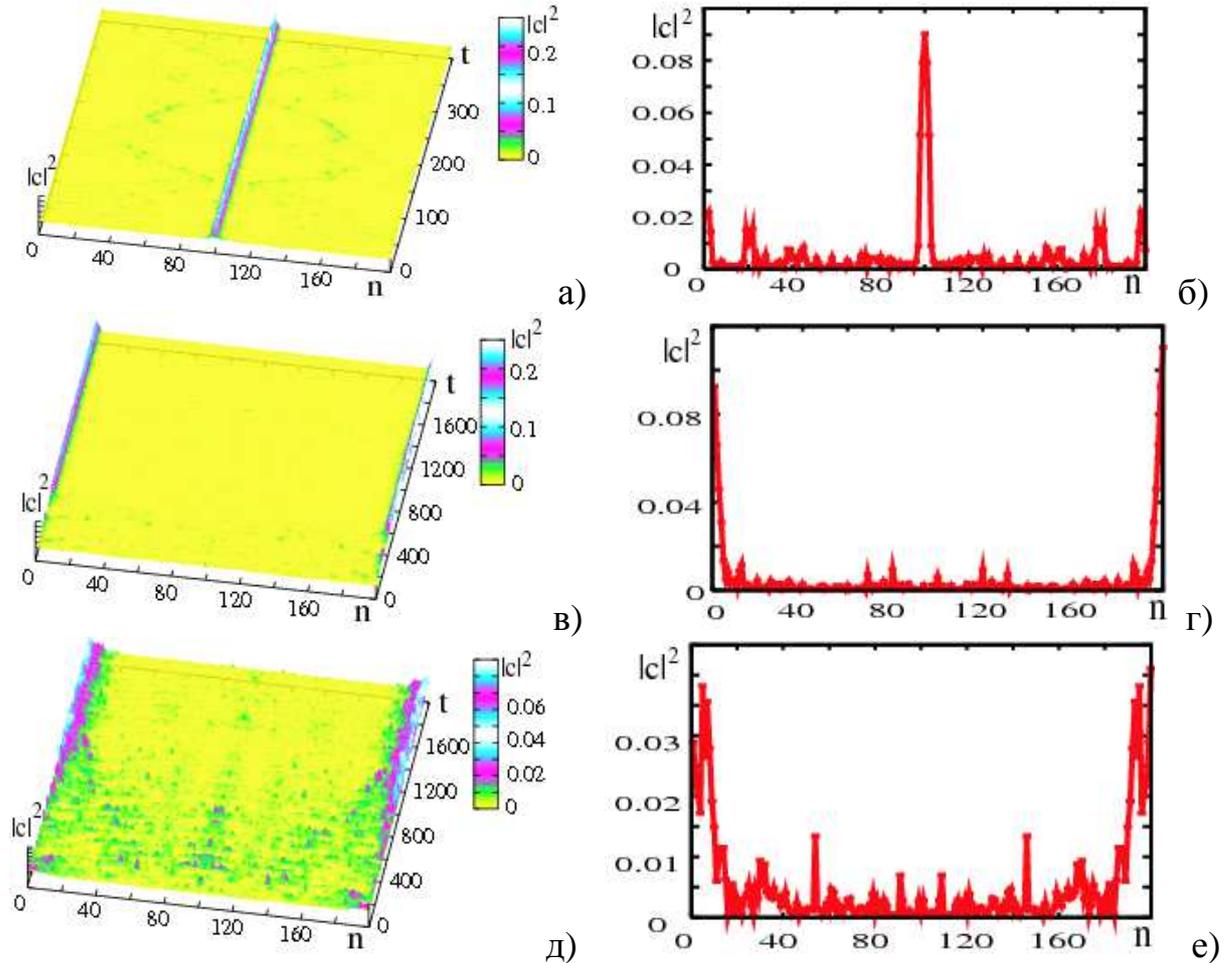

**Рис. 10.** Формирование полярона первоначально локализованным электроном при различных значениях параметра взаимодействия χ. В левой колонке отображена эволюция во времени распределения плотности вероятности обнаружить электрон в цепочке, в правой – его состояние в последний момент симуляции. $N = 200$, $\tau_e = 18$, $\gamma = 0.001$. (а, б) $\chi_h = 4$, $\chi_{el} = 48$; (в, г) $\chi_h = 2$, $\chi_{el} = 34$; (д, е) $\chi_h = 1$, $\chi_{el} = 17$.

Для исследования влияния бризера на полярон с целью определения возможности захвата полярона бризером или стимуляции движения полярона (без захвата), будем рассматривать систему с относительно большим параметром $\chi_h = 4$, в которой времена эволюции возмущений достаточно короткие (см. рис. 10,а и 10,б). В компьютерных экспериментах в решетке, в которой уже сформирован полярон, дополнительно возбуждался баббл (бризер) благодаря воздействию внешнего импульса (скорости $v_{in}$) на одну из частиц с номером $n_{in}$, причем величины $v_{in}$ и $n_{in}$ варьировались в широком диапазоне значений. Типичные результаты представлены на рис. 11 и 12. Прежде всего из них следует, что полярон никогда не захватывается бризером, поскольку структура полярона и бризера не соответствуют друг другу. Основной наблюдаемый эффект – это стимулированное возбуждаемым бризером движение полярона, даже если расстояние между бризером и поляроном достаточно большое (см. рисунки 11,в и 12). Можно предполагать, что полярон начинает двигаться в результате взаимодействия с фононами, возникающими в решетке при формировании бризера (баббла). Энергия фононов существенно меньше энергии бризера, однако достаточна чтобы стимулировать движение полярона. Полярон может двигаться как к бризеру (рис. 11), так и от него (рис. 12). В первом случае при встрече с бризером полярон может пересечь его, если бризер узкий (с малой энергией), как показано на рис. 11,а–в, или отразится от него, если бризер (баббл) мощный (рис. 11,г–е). При этом полярон может смещаться на большое расстояние, порядка 100–200 сайтов, двигаясь с почти постоянной скоростью. Таким образом, возбуждая в решетке бризер, можно стимулировать движение существующих в решетке поляронов, т. е. обеспечить транспорт заряда. Закономерности движения полярона в зависимости от параметров бризера (параметров внешнего воздействия) еще предстоит выяснить, и тогда можно будет уверенно говорить о возможности управлении характеристиками транспорта заряда в ДНК.

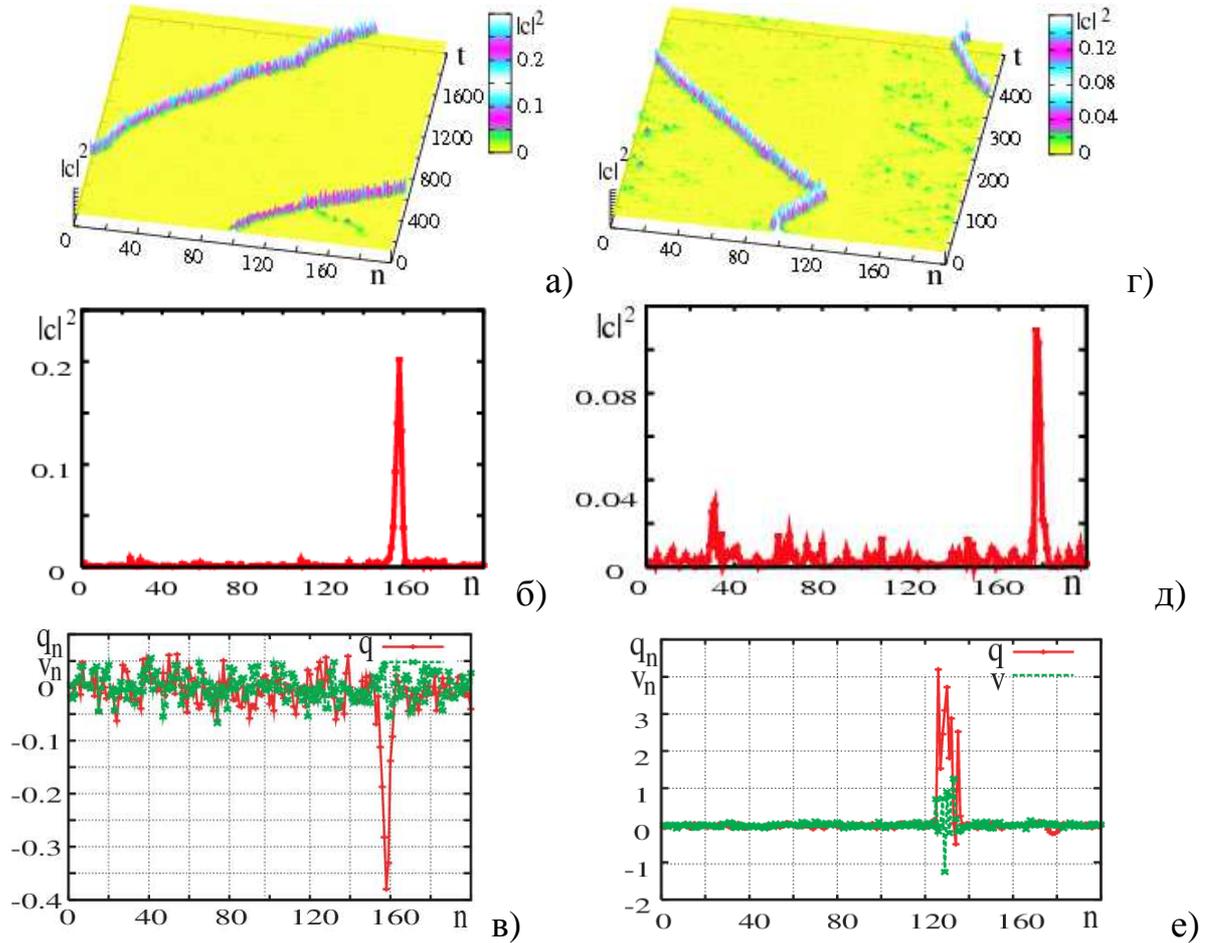

**Рис. 11.** Движение полярона в результате возбуждения бризера или баббла. Представлены картины эволюции во времени распределения плотности вероятности обнаружить электрон в цепочке, а), г) распределение плотности вероятности обнаружить электрон на *n*-ом сайте, б), д), и координат $q_n$ (красным) и скоростей частиц $v_n$ (зеленым), в), е) вдоль цепочки в последний момент симуляции. Рисунки в левой колонке относятся к случаю возбуждения бризера слабым внешним воздействием ($v_{in} = 0.1$, $n_{in} = 150$), справа – к случаю возбуждения баббла сильным внешним воздействием ($v_{in} = 5$, $n_{in} = 130$). $N = 200$, $\chi_h = 4$, $\tau_e = 18$, $\chi_{el} = 48$, $\gamma = 0.001$.

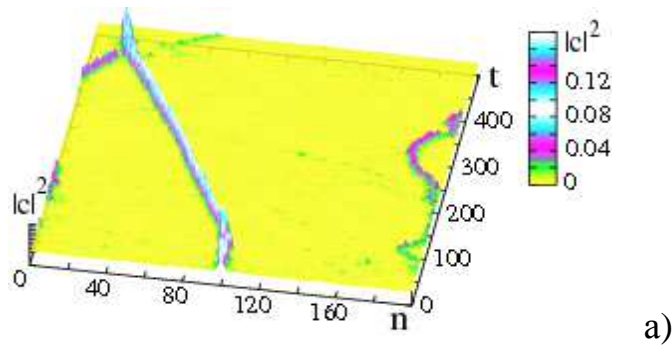

а)

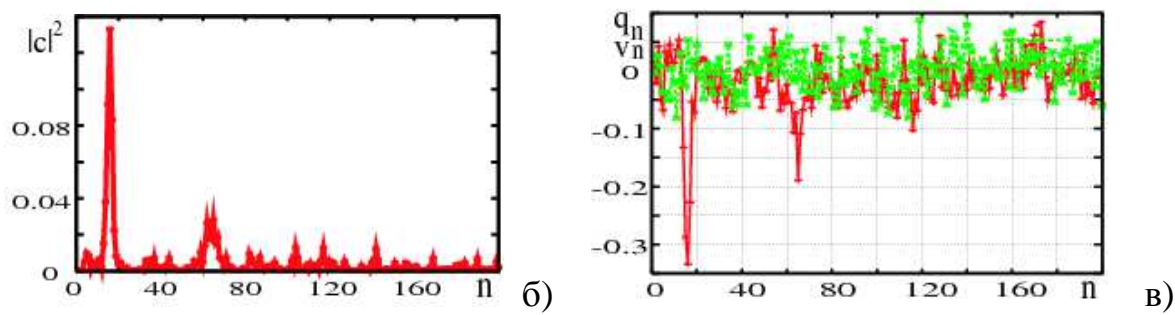

**Рис. 12.** Движение полярона в результате возбуждения бризера слабым внешним воздействием (от рис. 11 отличается тем, что первоначально полярон движется от возбужденного в решетке бризера.) Представлены картины эволюции во времени распределения плотности вероятности обнаружить электрон в цепочке, а), распределение плотности вероятности обнаружить электрон на n-ом сайте, б), и координат $q_n$ (красным) и скоростей частиц $v_n$ (зеленым), в), е) вдоль цепочки в последний момент симуляции. $N = 200$, $\chi_h = 4$, $\tau_e = 18$, $\chi_{el} = 48$, $\gamma = 0.001$, $v_{in} = 0.01$, $n_{in} = 150$.

## ОБСУЖДЕНИЕ РЕЗУЛЬТАТОВ

Показано, что при локальном внешнем интенсивном возбуждении решетки ДНК возможно образование бризеров на движущихся фронтах возникшего вследствие нелинейности решетки баббла. Эти бризеры в состоянии захватывать внешний электрон и перемещать его вдоль решетки. Характерное время существования связанного состояния «бризер–электрон» порядка 20–30 периодов осцилляций сайтов решетки. Характерная длина смещения ~ 20 сайтов. Максимальная вероятность обнаружить электрон в движущемся бризере ~ 0.2. Обнаружено, что отклонение частиц от равновесного состояния в образующемся баббле очень большое. Возможно, что оно будет корректироваться при использовании более точной модели решетки, в частности описанием взаимодействия сайтов решетки в рамках модели Пейрарда–Бишопа–Доксуа.

Продемонстрировано, что если в решетке существует неподвижный полярон, а затем возбуждается баббл, даже не очень мощный, то в результате взаимодействия полярона с фононами, излученными при формировании баббла, полярон приобретает способность двигаться с достаточно высокой скоростью, при этом первоначальное направление движения зависит, по-видимому, от фазы фононных волн, достигающих полярона. Движущийся полярон может пересечь область низкоэнергетического баббла или отразиться от мощного баббла. Характерная длина траектории, т. е. расстояния, на которое переносится заряд, может быть очень большой, до 100–200 сайтов. Можно надеяться, что более детальное изучение динамики такого взаимодействия даст возможность управления характеристиками транспорта заряда в ДНК, обеспечиваемого таким механизмом.

Для экспериментальной реализации обнаруженных явлений можно предполагать использование атомного сканирующего зондового микроскопа, игла которого подводится к молекуле ДНК и одновременно инжектирует в неё заряд, что приводит к локальному разогреву молекулы и образованию в этом месте инжекции заряда солитона или баббла, как рассмотрено в [32].

Можно предполагать также, что игла АСМ на одном конце ДНК коротким импульсом или набором импульсов возбуждает в молекуле локализованные возмущения, солитоны или бризеры (или наборы их), которые распространяясь вдоль молекулы производят захват поляронов или стимулируют их движение, обеспечиваю их транспорт по ДНК.

Рассмотренный в работе эффект возбуждения баббла периодическим воздействием на короткий участок ДНК требует для осуществления высокочастотного воздействия с частотой $\omega \sim 10^{12}$ сек$^{-1}$. В настоящее время генерация воздействий на молекулу такой частоты представляет собой проблему, т.к. в реальных силовых микроскопах максимальная частота не превосходит $10^5$–$10^6$ Гц [33], однако развитие терагерцовых технологий дает надежду на решение этой задачи.